\documentclass[12pt]{article}

\usepackage{epsf}
\usepackage{latexsym,amssymb,euscript}
\usepackage[dvips]{graphicx}
\usepackage{amsmath}
\usepackage[nolists,nomarkers]{endfloat}

\begin{document}

\begin{center}
{\Large \textbf{Critical behavior of $3D$ $Z(N)$ lattice gauge theories 
at zero temperature}}
\end{center}

\vskip 0.3cm
\centerline{O.~Borisenko$^{1\dagger}$, V.~Chelnokov$^{1*}$, 
G.~Cortese$^{2\dagger\dagger}$, M.~Gravina$^{3\ddagger}$, A.~Papa$^{3\P}$, 
I.~Surzhikov$^{1**}$}

\vskip 0.6cm

\centerline{${}^1$ \sl Bogolyubov Institute for Theoretical Physics,}
\centerline{\sl National Academy of Sciences of Ukraine,}
\centerline{\sl 03680 Kiev, Ukraine}

\vskip 0.2cm

\centerline{${}^2$ \sl Instituto de F\'{\i}sica Te\'orica UAM/CSIC,}
\centerline{\sl Cantoblanco, E-28049 Madrid, Spain}
\centerline{\sl and Departamento de F\'{\i}sica Te\'orica,}
\centerline{\sl Universidad de Zaragoza, E-50009 Zaragoza, Spain}

\vskip 0.2cm

\centerline{${}^3$ \sl Dipartimento di Fisica, Universit\`a della 
Calabria,}
\centerline{\sl and Istituto Nazionale di Fisica Nucleare, 
Gruppo collegato di Cosenza}
\centerline{\sl I-87036 Arcavacata di Rende, Cosenza, Italy}

\vskip 0.6cm

\begin{abstract}
Three-dimensional $Z(N)$ lattice gauge theories at zero temperature are 
studied for various values of $N$. Using a modified phenomenological 
renormalization group, we explore the critical behavior of the generalized 
$Z(N)$ model for $N=2,3,4,5,6,8$.  
Numerical computations are used to simulate vector models for 
$N=2,3,4,5,6,8,13,20$ for lattices with linear extension up to $L=96$. We 
locate the critical points of phase transitions and establish their scaling 
with $N$. The values of the critical indices indicate that the models with 
$N>4$ belong to the universality class of the three-dimensional $XY$ model. 
However, the exponent $\alpha$ derived from the heat capacity is consistent 
with the Ising universality class. We discuss a possible resolution of this 
puzzle. We also demonstrate the existence of a rotationally symmetric region 
within the ordered phase for all $N\geq 5$ at least in the finite volume. 
\end{abstract}

\vfill
\hrule
\vspace{0.3cm}
{\it e-mail addresses}:

$^\dagger$oleg@bitp.kiev.ua, \ \ $^*$chelnokov@bitp.kiev.ua,
\ \ $^{\dagger\dagger}$cortese@unizar.es,

$^{\ddagger}$gravina@cs.infn.it, \ \ 
$^{\P}$papa@cs.infn.it, \ \ $^{**}$i\_van\_go@inbox.ru

\newpage 

\section{Introduction}

Models possessing global and/or local discrete $Z(N)$ symmetry 
play an important role in many branches of physics ranging from 
the solid state physics to the description of the universality features 
of the deconfining transition in $SU(N)$ gauge theories. 
In this paper we are interested in the $Z(N)$ lattice gauge theory (LGT) at 
zero temperature. Assigning the gauge fields $s_n(x)=0,1,\cdots,N-1$ to the 
links of a simple hypercubic lattice, the most general action of the 
isotropic $Z(N)$ LGT can be written as 
\begin{equation} 
S_{\rm gauge}  =  \frac{1}{2} \  \sum_x \sum_{n<m} \ \sum_{k=1}^{N-1} \beta_k
\exp \left( \frac{2 \pi i k}{N} \left(s_n(x) + s_m(x+e_n) 
-s_n(x+e_m) - s_m(x) \right) \right) \ ,
\label{action_gauge}
\end{equation}
where $e_n$, $n=1,2,3$, denotes the unit vector in the $n$-th direction. 
Similarly, the most general action of the $Z(N)$ spin model is given by  
\begin{equation} 
S_{\rm spin} \ = \ \frac{1}{2} \ \sum_x \sum_{n} \ \sum_{k=1}^{N-1} \beta_k
\exp \left( \frac{2 \pi i k}{N} \left(s(x) - s(x+e_n) \right) \right) \ .
\label{action_spin}
\end{equation}
In both cases we used the convention
\begin{equation} 
\beta_k \ = \ \beta_{-k} \ = \ \beta_{k+N}  \ > \ 0 \ .
\label{beta_convention}
\end{equation}
The standard Potts model corresponds to the choice when all $\beta_k$ are 
equal. Then, the sum over $k$ reduces to a delta-function on the $Z(N)$ group. 
The conventional vector model corresponds to $\beta_k=0$ for all $k\ne 1,N-1$. 
For $N=2,3$ the Potts and vector models are equivalent. 

Two-dimensional (2D) standard and vector $Z(N)$ LGTs are exactly solvable both 
in the finite volume and in the thermodynamic limit. They exhibit no phase 
transition at any finite value of the coupling constant $\beta$. 
In particular, the rectangular $R\times T$ Wilson loop in the representation 
$k$ obeys the area law 
\begin{equation*} 
\langle  W_k(S) \rangle  \ = \ \exp \left (-\sigma_k (N) \ R \ T   \right ) \ ,
\end{equation*}
thus implying a permanent confinement of static charges. For example, the 
string tension of the vector model in the thermodynamic limit reads 
\begin{eqnarray*} 
\sigma_k(N)  \ = \ \ln \left [ \frac{C_0(N,\beta)}{C_k(N,\beta)} \right ] \ > 
0 \ \ , \ \ C_k(N,\beta) \ = \ \sum_{r=-\infty}^{\infty} \ I_{Nr+k}(\beta) \ . 
\end{eqnarray*}
Here, $I_k(x)$ is the modified Bessel function. 

No exact solution has been found for any $Z(N)$ model in $3D$, where the phase 
structure becomes highly non-trivial. While the phase structure of the general 
model defined by~(\ref{action_gauge}) remains unknown, it is well established 
that Potts models and vector models with only $\beta_1$ non-vanishing have one 
phase transition from a confining phase to a phase with vanishing string 
tension~\cite{horn,ukawa,savit}. 
Via duality, $Z(N)$ gauge models can be exactly related to $3D$ $Z(N)$ spin 
models. In particular, a Potts gauge theory is mapped to a Potts spin model, 
and such a relation allows to establish the order of the phase 
transition. Hence, Potts LGTs with $N=2$ have second order phase transition, 
while for $N\geq 3$ one finds a first order phase transition. 
Since $Z(2)$ LGT is equivalent to the Ising model, its critical behavior is 
well known (see Refs.~\cite{caselle} and references therein). Generally, the 
$Z(N)$ global symmetry of the finite-temperature $4D$ $SU(N)$ gauge theory 
motivated thorough investigations, both analytical and numerical,  
of the $3D$ spin models, especially for $N=2,3$~\cite{gavai,fukugita} (for 
more recent studies, see~\cite{bazavov_z3} and references therein). 
The $3D$ Potts models for $N>3$ have been simulated in~\cite{bazavov} 
and studied by means of the high-temperature expansion in~\cite{janke}. 

Surprisingly, much less is known about the critical behavior of $Z(N)$ vector 
LGTs when $N>4$. They have been studied numerically in~\cite{bhanot} up 
to $N=20$ on symmetric lattices with size $L\in [4-16]$. It was confirmed that 
zero-temperature models possess a single phase transition which disappears in 
the limit $N\to\infty$. A scaling formula proposed in~\cite{bhanot} shows that 
the critical coupling diverges like $N^2$ for large $N$. 
Thus, the $U(1)$ LGT has a single confined phase in 
agreement with theoretical results~\cite{3d_u1}. 
We are not aware, however, of any detailed study of the critical behavior 
of the vector models with $N\geq 4$ in the vicinity of this single phase 
transition. Slightly more is known about the critical properties of $Z(N)$ 
vector spin models. In particular, it has been suggested that all vector spin 
models exhibit a single second order phase transition~\cite{scholten}. 
An especially detailed study was performed on the $Z(6)$ model, because the
$Z(6)$ global symmetry appears as an effective symmetry of the $Z(3)$ 
antiferromagnetic Potts model~\cite{z6_vector_mc,z6_vector_rg}. The computed 
critical indices suggest that the $Z(6)$ vector model belongs to the 
universality class of the $3D$ $XY$ model. An interesting feature of the $Z(6)$
model and, possibly, of all vector models with $N>4$, is the appearance of an
intermediate rotationally symmetric region below the critical temperature of the 
second order phase transition. The mass gap was however found to be rather 
small, but non-vanishing in this region~\cite{z6_vector_mc}. 
Combined with a renormalization group (RG) study, the analysis 
concluded that this intermediate region presents a crossover to a
low-temperature massive phase, where the discreteness of $Z(6)$ plays an 
essential role~\cite{z6_vector_rg}.      

The main goal of the present work is to fill the gap in our knowledge about 
the critical behavior of the $3D$ $Z(N)$ LGTs. Another motivation comes from 
our recent studies of the deconfinement transition in the $Z(N)$ vector LGT 
for $N>4$ at finite temperatures~\cite{3d_zn_strcoupl,lat_12,ZN_fin_T}. The 
major findings of these papers was the demonstration of two phase transitions 
of the Berezinskii-Kosterlitz-Thouless type and the existence of an 
intermediate massless phase. The critical indices at these transitions have 
been found to coincide with the indices of the $2D$ vector spin models. An 
interesting question then arises regarding the construction of the continuum 
limit of the finite-temperature models in the vicinity of the critical points. 
For this to accomplish it might be useful, and even necessary, to know the 
scaling of quantities such as string tension, correlation length, 
{\it etc.} near the critical points of the corresponding zero temperature 
models.  

In this work we are going to: 
\begin{itemize} 

\item 
perform an analytical study of the general $3D$ $Z(N)$ LGT for various $N$ 
using a ``phenomenological'' RG; 

\item 
locate critical points of the vector models via Monte Carlo simulations of 
the dual of the $Z(N)$ LGT and determine their scaling with $N$; 

\item 
compute some critical indices and establish the universality class of the 
models;

\item 
illustrate the existence of a rotationally symmetric region within the 
ordered phase for all $N\geq 5$ at least in the finite volume. 

\end{itemize}

This paper is organized as follows. In Section~2 we formulate our model and 
recall the exact duality relation with a generalized $3D$ $Z(N)$ spin model. 
Section~3 is devoted to a RG study of the models. Within a modified version 
of the phenomenological RG, we explore the space of coupling constants, find 
fixed points and calculate the critical index $\nu$.
In Section~4 we present the setup of Monte Carlo simulations, define the 
observables used in this work and present the numerical results. In particular,
we locate the position of critical points and compute various critical indices 
at these points. As a cross-check, we also simulated $Z(N=2,3)$ models for 
which high-precision results. The same Section deals also with the computation 
of the average action and the heat capacity in the vicinity of critical points 
and the derivation of the index $\alpha$ from a finite size scaling (FSS) 
analysis of the heat capacity. In Section~5 we discuss some findings regarding 
the symmetric region below the critical point. All results are finally 
summarized in Section~6.

\section{Relation of the $3D$ $Z(N)$ LGT to a generalized $3D$ $Z(N)$ spin 
model} 

We work on a $3D$ lattice $\Lambda = L^3$ with linear extension $L$; 
$\vec{x}=(x_1,x_2,x_3)$, where $x_i\in [0,L-1]$ denote the sites of the 
lattice and $e_n$, $n=1,2,3$, denotes the unit vector in the $n$-th direction.
Periodic boundary conditions (BC) on gauge fields are imposed in all 
directions. We introduce conventional plaquette angles $s(p)$ as
\begin{equation}
s(p) \ = \ s_n(x) + s_m(x+e_n) - s_n(x+e_m) - s_m(x) \ .
\label{plaqangle}
\end{equation}
The $3D$ $Z(N)$ LGT on an isotropic lattice can generally be defined as 
\begin{equation}
Z(\Lambda ;\{ \beta_k \};N) \ = \  \prod_{l\in \Lambda}
\left ( \frac{1}{N} \sum_{s(l)=0}^{N-1} \right ) \ \prod_{p} Q(s(p)) \ .
\label{PTdef}
\end{equation}
The most general $Z(N)$-invariant Boltzmann weight is
\begin{equation}
Q(s) \ = \
\exp \left [ \frac{1}{2}\sum_{k=1}^{N-1} \beta_k 
\exp \left ( \frac{2\pi i k}{N}s  \right ) \right ] \ .
\label{Qpgen}
\end{equation}
The $U(1)$ LGT is defined as the limit $N\to\infty$ of the above expressions.

To study the phase structure of $3D$ $Z(N)$ LGTs one can map the gauge model 
to a generalized $3D$ $Z(N)$ spin model with the action given by 
Eq.~(\ref{action_spin}). 
The relation between spin $\beta_k^s$ and gauge $\beta_k^g$ couplings can be 
computed exactly and reads  
\begin{equation}
\beta_k^s \ =\ \frac{1}{N} \sum_{p = 0}^{N - 1} \ln \left [ \frac{Q_d(p)}
{Q_d(0)} \right ] \  \cos \left(\frac{2 \pi p k}{N} \right) \ , 
\label{couplings}
\end{equation}
where the dual Boltzmann weight is defined as 
\begin{equation}
Q_d(s) \ = \ \sum_{p = 0}^{N - 1} \ Q(p) \ \cos \left(\frac{2 \pi p s}{N} 
\right) \ .
\label{Qdual}
\end{equation} 
In what follows we are going to simulate the $Z(N)$ vector LGT. In this case 
we use $\beta_1=\beta_{N-1}=\beta$ and the dual weight becomes 
\begin{equation}
Q_d(s) \ = \ \sum_{r=-\infty}^{\infty} \ I_{Nr+s}(\beta) \ .
\label{Qdual_vector}
\end{equation}
An example of the explicit relations between couplings in this case together 
with some further important comments on the relations can be found 
in~\cite{ZN_fin_T}.

With the goal of performing RG transformations, it is somewhat more convenient 
to use a different but equivalent representation for the Boltzmann weight, 
namely 
\begin{equation}
Q\left[ \{ t_k \}; s \right ] \ = \ \sum_{k=0}^{N-1} \ t_k \ 
\exp\left[ \frac{2\pi i}{N} \ k \ s \right ] \ .
\label{Bweight1}
\end{equation}
The set of coupling constants $\{ t_k \}$ can be chosen to satisfy 
\begin{equation*}
t_0=1 \ , \ 0 \leq t_k \leq 1 \ , \ t_k=t_{-k}=t_{k+N} \ .
\end{equation*}
The coupling constants $t_k$ and $\beta_k$ can be connected to each other via 
the Fourier transform on the $Z(N)$ group. The dual of the partition function 
(\ref{PTdef}) can then be presented as 
\begin{equation}
Z(\Lambda ;\{ t_k \};N) \ = \  \sum_{j_n=0}^{N-1} \ 
\prod_{x\in \Lambda} \ 
\left ( \frac{1}{N} \sum_{s(x)=0}^{N-1} \right ) \ \prod_{x,n} \ t_{s(x)-s(x+e_n)+\eta_n} \ .
\label{PT_dual}
\end{equation}
Here, the summations over $j_n, n=1,2,3$, enforce the global Bianchi 
constraints due to the periodic BC and $\eta_n=j_n$ on a set of links dual 
to any fixed closed surface wrapping the original lattice in the 
directions perpendicular to $n$, otherwise $\eta_n=0$.

\section{Results of the RG study} 

As a first step we consider the general $3D$ $Z(N)$ LGT with the Boltzmann 
weight (\ref{Bweight1}) and study its phase structure with the help of 
the phenomenological renormalization group (PH RG)~\cite{nightingale}. For the 
$3D$ Ising model this RG was used in~\cite{yurishchev}. As is well known, 
the PH RG gives in many cases not only the qualitatively correct phase diagram 
of a model, but also a good quantitative approximation for the observables 
near the critical points, and this approximation can be systematically 
improved. We follow the general strategy described in the original 
papers~\cite{nightingale} and only briefly outline our modifications 
(for a detailed account of these modifications, together with application 
to other models, see~\cite{sdarg}). As is common for this type of RG,
we preserve the mass gap during the RG steps. We proceed as follows. 

\begin{enumerate}

\item  As a starting point we always use the dual formulations, {\it i.e.} 
Eq.~(\ref{PT_dual}) in the present case. 

\item As a simplest $3D$ strip, we take a lattice 
$\Lambda_s=(2\times 2\times L)$ with periodic BC in both transverse directions 
and periodic or free BC in the longitudinal direction. Correlation functions 
are computed on $\Lambda_s$ and on a one-dimensional lattice for all 
representations. 

\item The requirement of mass gap preservation leads to equations for 
the fixed points 
\begin{equation}
t_j^{(1)} \ = \ \left [ \frac{\lambda_j(\{ t_k \})}{\lambda_0(\{ t_k \})} 
\right ]^2 \ ,
\label{RRM2}
\end{equation}
where $\lambda_j(\{ t_k \})$ is the eigenvalue of the transfer matrix 
corresponding to the correlation function in the representation $j$. This set 
of equations is treated as the recursion relations for the renormalized 
coupling constants. 

\item The summation over the spins lying in one plane perpendicular to the 
longitudinal direction introduces all possible interactions between the 
remaining spins. The transfer matrix is constructed for the evolution of all 
independent couplings of this general interaction. This substantially reduces 
the size of the matrix.  

\item Combining the PH RG in the form described above with the cluster 
decimation approximation of Ref.~\cite{CDA}, we can construct new partition 
and correlation functions on a decimated lattice with double lattice spacing. 
The next iteration is performed with newly computed constants $t_j^{(1)}$.   

\end{enumerate}

In the framework of this approach we have explored the phase structure of the 
generalized $Z(N)$ model. Only in the case of the standard Potts model we 
can restrict ourselves to one iteration, since the RG steps do not generate 
new interactions. For the general case one should perform many iterations to 
locate precisely the critical points. The critical indices are calculated 
in the fixed point of the iterations where the critical points of the models 
of a given universality class are flowing to. Thorough discussion of the general 
case will be given in~\cite{sdarg}. Here we report on the results for 
the vector LGTs relevant for this paper. 

\begin{table}[ht]
\caption{$3D$ $Z(N)$ standard Potts models: $\beta_{\rm c}$ from the PH RG 
(column two) and from the Monte Carlo simulations of Ref.~\cite{bazavov} 
(column three).
\newline
$3D$ $Z(N)$ vector models: $\beta_{\rm c}$ from the PH RG (column~4) and from 
the Monte Carlo simulations of this work (column five); critical index $\nu$
from the PH RG.
}
\begin{center}
\begin{tabular}{|c|c|c|c|c|c|}
\hline
    & \multicolumn{2}{|c|}{Potts model} & \multicolumn{3}{|c|}{Vector model}\\
\cline{2-6}
$N$ & $\beta_{\rm c}$ & $\beta_{\rm c}^{\rm MC}$ & $\beta_{\rm c}$ 
& $\beta_{\rm c}^{\rm MC}$ & $\nu$ \\ 
\hline
2  & 0.77706 & 0.761414(2) & 0.77706 & 0.761395(4) & 0.616656 \\
3  & 1.17186 & 1.084314(8) & 1.17186 & 1.0844(2)   & -        \\
4  & 1.34363 & 1.288239(5) & 1.55411 & 1.52276(4)  & 0.616657 \\
5  & 1.50331 & 1.438361(4) & 2.17896 & 2.17961(10) & 0.692226 \\
6  & 1.62881 & 1.557385(4) & 2.99296 & 3.00683(7)  & 0.699208 \\
8  & 1.81941 & 1.740360(6) & 5.09472 & 5.12829(13) & 0.699583 \\
13 & -       & -           & -       & 13.1077(3)  & -        \\
20 & -       & -           & -       & 30.6729(5)  & -        \\
\hline
\end{tabular}     
\end{center}
\label{rg_results}
\end{table} 

 In Table~\ref{rg_results} we present the estimates of the critical points
$\beta_{\rm c}$, both in standard Potts models and in vector models, coming 
from the PH RG, and compare them with the results of Monte Carlo
numerical simulations. In the case of standard Potts models, these simulations
were performed in Ref.~\cite{bazavov}, while for the vector models, they
were carried out in this paper. One can see that the PH RG based on the
smallest possible strip and combined with the cluster decimation approximation 
gives quite accurate predictions both for the critical coupling and, as we will
see later on, also for the index $\nu$.  

\section{Numerical results}

\subsection{Setup of the Monte Carlo simulation} 

The model under exam in this work is described by the action given in 
Eqs.~(\ref{PTdef}) and~(\ref{Qpgen}), with couplings $\beta_1=\beta_{N-1}\equiv
\beta$, $\beta_n=0$, $n=2,\ldots,N-2$. To study the phase transitions 
it turns out to be more convenient to simulate the dual spin model, 
whose action is given in~(\ref{action_spin}) with the coupling constants 
computed according to Eq.~(\ref{couplings}).  
Simulations were performed by means of a cluster algorithm on symmetric 
lattices $L^3$ with periodic BC and $L$ in the range 8 -- 96.
For each Monte Carlo run the typical number of generated configurations 
was $2.5 \cdot 10^6$, the first $10^5$ of them being discarded to ensure 
thermalization. Measurements were taken after every 10 updatings and error 
bars were estimated by the jackknife method combined with binning.

We considered the following observables:
\begin{itemize}
\item complex magnetization $M_L = |M_L| e^{i \psi}$,
\begin{equation}
\label{complex_magnetization}
M_L \ =\  \sum_{x \in \Lambda} \exp \left( \frac{2 \pi i}{N} s(x) \right) \;;
\end{equation}

\item population $S_L$,
\begin{equation}
\label{population}
S_L \ =\  \frac{N}{N - 1} \left(\frac{\max_{i = 0, N - 1} n_i} {L^3} 
- \frac{1}{N} \right)\;,  
\end{equation}
where $n_i$ is number of $s(x)$ equal to $i$;

\item real part of the rotated magnetization $M_R = |M_L| \cos(N \psi)$
and normalized rotated magnetization $m_\psi = \cos(N \psi)$;

\item susceptibilities of $M_L$, $S_L$ and $M_R$:  
$\chi_L^{(M)}$, $\chi_L^{(S)}$, $\chi_L^{(M_R)}$
\begin{equation}
\label{susceptibilities}
\chi_L^{(\mathbf\cdot)} \ =\  L^2  \left(\left< \mathbf\cdot^2 \right> 
- \left< \mathbf\cdot \right>^2 \right)\;;
\end{equation}

\item Binder cumulants $U_L^{(M)}$ and $B_4^{(M_R)}$,
\begin{eqnarray}
U_L^{(M)}&\ =\ &1 - \frac{\left\langle \left| M_L \right| ^ 4 
\right\rangle}{3 \left\langle \left| M_L \right| ^ 2 \right\rangle^2}\;, 
\nonumber \\
\label{binderU}
B_4^{(M_R)}&\ =\ & \frac{\left\langle \left| M_R 
- \left\langle M_R \right\rangle \right| ^ 4 \right\rangle}
{\left\langle \left| M_R - \left\langle M_R \right\rangle \right| ^ 2 
\right\rangle ^ 2 } \ . 
\label{binderBMR}
\end{eqnarray}

\end{itemize} 
We computed also the average action and the heat capacity in the vicinity of 
the critical points. 

\subsection{Critical couplings and their scaling with $N$}

We obtained the critical couplings using the Binder cumulant crossing method
described in~\cite{3dxy_univ}. In particular, we computed by Monte Carlo
simulations the Binder cumulant $U_L^{(M)}$ and its first three derivatives
with respect to $\beta$ for the different lattice sizes, thus allowing 
to build the function $U_L^{(M)}(\beta)$ in the region near the transition. Then, 
we looked for the value of $\beta$ at which the curves $U_L^{(M)}(\beta)$
related to the different lattice sizes $L$ ``intersect''. In fact, 
the critical coupling $\beta_{\rm c}$ was estimated as the value of 
$\beta$ at which $U_L^{(M)}(\beta)$ exhibits the least dispersion over 
lattice sizes ranging from $L=16$ to $L=96$.

The values of $\beta_{\rm c}$ are quoted in the column five of 
Table~\ref{rg_results}. The error bars take into account some of the systematic
effects, the main one being the dependence of the estimated value of 
$\beta_{\rm c}$ on the set of lattice sizes considered in the analysis. 
Another, less relevant, source of systematics is the uncertainty in the
analytic dependence of $U_L^{(M)}$ on $\beta$ near the critical point.

\begin{figure}
\center{\includegraphics[width=0.75\textwidth]{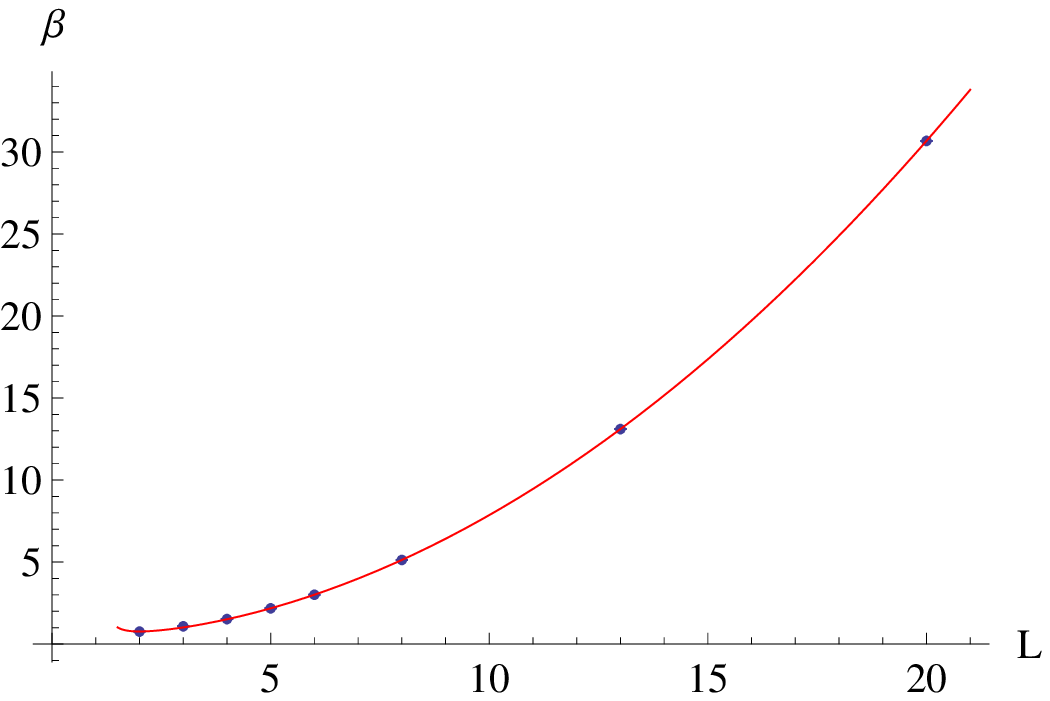}}
\caption{Critical couplings of $3D$ $Z(N)$ vector models plotted versus 
$N$ (dots) together with the fitting curve given in~(\ref{fit_betac})
(solid line). The error bars on the critical couplings are smaller than 
symbols.}
\label{critical_coupling_fit}
\end{figure}

In~\cite{bhanot} the dependence of the $3D$ $Z(N)$ critical couplings on 
$N$ was described as 
\begin{equation}
\beta_{\rm_c} = \frac{1.5}{1-\cos\left(\frac{2 \pi}{N}\right)}\ .
\label{fit_betac_old}
\end{equation}
With our new data we checked if the value 1.5 is exact and provided the 
next-order correction to it. Using different forms for the next-order 
corrections, we found that our data exclude a correction of 
${\cal O}(N^0)$, but reveal corrections of the order $1/N^2$, 
which we take in the form $C\left({1-\cos\left(\frac{2\pi}{N}\right)}\right)$.
The critical coupling values for the $3D$ $Z(N>4)$ vector models were then
fitted with the formula 
\begin{equation}
\beta_{\rm c} = \frac{A}{1-\cos\left(\frac{2 \pi}{N}\right)} 
+ C \left({1-\cos\left(\frac{2 \pi}{N}\right)}\right) \ ,
\label{fit_betac}
\end{equation}
giving the following results:
$A=1.50122(7)$, $C=0.0096(5)$, $\chi^2/{\rm d.o.f.} = 13.1$ (see 
Fig~\ref{critical_coupling_fit}). Despite the large $\chi^2$, 
probably due to the underestimation of the error bars of critical
couplings, the proposed function nicely interpolates data over a large
interval of values of $N$. 

\subsection{Critical indices and hyperscaling relation}

The procedure to determine the critical index $\nu$ is also inspired 
by Ref.~\cite{3dxy_univ}: for each lattice size $L$ the known function
$U_L^{(M)}(\beta)$ is used to determine $d U_L^{(M)}(\beta)/d\beta$; from this, the
derivative of $U_L^{(M)}$ with respect to the rescaled coupling 
$x = (\beta - \beta_{\rm c}) L^{1/\nu}$ can be calculated,
\begin{equation}
\frac{d U_L^{(M)}} {d x} = \frac{d U_L^{(M)}}{d \beta} L^{1/\nu}\;.
\label{nu_determ}
\end{equation}
The best estimate of $\nu$ is found by minimizing the deviation of 
$d U_L^{(M)}/d x$ with respect to a constant value. The minimization can be done
at $\beta_{\rm c}$ or at any other value $\beta_{\rm f} \approx \beta_{\rm c}$ 
defined as the point where $U_L^{(M)}$ on a given lattice becomes equal to some 
fixed value. The resulting values for $\nu$, summarized in 
Table~\ref{indices_nu_zero}, do not differ within error bars.

\begin{table}
\caption{Critical index $\nu$ of $3D$ $Z(N)$ vector models determined 
by the fit procedure described in the text for different choices of the 
minimum lattice size $L_{\rm min}$; the $\chi^2$ given in the last column 
is the reduced one }
\begin{center}
\setlength{\tabcolsep}{4.5pt}
\begin{tabular}{|c|c|c|c|}
\hline
$N$ & $L_{\min}$ & $\nu$  & $\chi^2_{\nu}$ \\
\hline
 2 & 8  & 0.6253(5) & 2.53 \\
   & 16 & 0.6280(7) & 1.21 \\
   & 24 & 0.6306(8) & 0.90 \\
\hline
 4 & 8  & 0.62661(11) & 1.96 \\
   & 16 & 0.62793(12) & 1.93 \\
   & 24 & 0.62933(12) & 1.07 \\
\hline
 5 & 8  & 0.6675(5) & 1.49 \\
   & 16 & 0.6698(4) & 0.94 \\
   & 24 & 0.6681(8) & 1.13 \\
\hline
 6 & 8  & 0.6687(12) & 4.12 \\
   & 16 & 0.6739(10) & 1.53 \\
   & 24 & 0.6756(17) & 2.10 \\
\hline
\end{tabular}
\hspace{1cm}
\setlength{\tabcolsep}{4.5pt}
\begin{tabular}{|c|c|c|c|}
\hline
$N$ & $L_{\min}$ & $\nu$  & $\chi^2_{\nu}$ \\
\hline
 8 & 8  & 0.6678(5) & 5.41 \\
   & 16 & 0.6720(4) & 1.96 \\
   & 24 & 0.6748(2) & 1.43 \\
\hline
 13& 8  & 0.6670(7)  & 3.68 \\
   & 16 & 0.6709(9)  & 2.42 \\
   & 24 & 0.6723(17) & 2.67 \\
\hline
 20& 8  & 0.6689(7) & 4.08 \\
   & 16 & 0.6730(4) & 1.04 \\
   & 24 & 0.6739(7) & 1.36 \\
\hline
\end{tabular}
\end{center}
\label{indices_nu_zero}
\end{table}

The critical indices $\beta/\nu$ and $\gamma/\nu$ can be extracted from 
the FSS analysis of the magnetization $M_L$ and its
susceptibility $\chi_L^{(M)}$, according to the following fitting functions,
\begin{eqnarray}
M_L &=& A_1 L^{-\beta/\nu} (1 + B_1 L^{-\delta}) \ , \nonumber \\
\chi_{M_L} &=& A_2 L^{\gamma/\nu} (1 + B_2 L^{-\delta}) \ ,
\label{fit_def}
\end{eqnarray}
which include also the first subleading corrections. The critical index
$\eta$ will then be given by $2-\gamma/\nu$ and the hyperscaling relation
$d=2\beta/\nu+\gamma/\nu$ must be satisfied with $d=3$.

In Tables~\ref{indices_part1} and~\ref{indices_part2} we summarize
the results for the critical indices, when the FSS analysis
is performed at $\beta$ fixed at the central value of the determination
of $\beta_{\rm c}$. Results are given for the cases when the subleading 
corrections, depending on the exponent~$\delta$, are included in neither fit, 
in both fits or only in one of them. When considered, the exponent $\delta$ 
has been fixed to the value $0.53/\nu$, as in the $3D$ $XY$ model. 
Varying $\delta$ in a wide interval does not give any significant change in 
the results.

In Tables~\ref{indices_0_method},~\ref{indices_1_method},~\ref{indices_2_method}
we summarize the values of the critical indices obtained by two
alternative methods: (i) performing the fit with the functions given
in~(\ref{fit_def}) not at $\beta_{\rm c}$, but at the pseudocritical
$\beta_{\rm pc}$ which maximizes the susceptibility $\chi_L^{(M)}$ 
(column two); (ii) performing the fit at a $\beta_{\rm f}$ defined as the 
point where $U_L^{(M)}$ on a given lattice becomes equal to some fixed value 
${U_{L,{\rm f}}^{(M)}}$; as these fixed values, we chose 
${U_{L,{\rm crit}}^{(M)}}$ -- our estimate for the intersection point of 
the $U_L^{(M)}$ curves at different lattice sizes (column three), then
a slightly larger value than this (column four) and slightly smaller 
value (column five) (for $N>4$, ${U_{L, {\rm f}}^{(M)}}=0.588$ in column three;
${U_{L,{\rm f}}^{(M)}}=0.60$ in column four; ${U_{L,{\rm f}}^{(M)}}= 0.57$ 
in column five); in all cases, the fit was done including data from lattice 
sizes with $L_{\rm min}=24$. 

\begin{table}
\caption{Critical indices $\beta/\nu$ and $\gamma/\nu$ of $3D$ $Z(N)$ vector 
models with $N=2,4,5,6$, determined by the fits in Eqs.~(\ref{fit_def}),
for different choices of the minimum lattice size $L_{\rm min}$. 
The $\chi^2$ of the two fits, given in columns four and six, are 
the reduced one. Column seven gives the dimension $d$ derived from the
hyperscaling relation $d=2\beta/\nu+\gamma/\nu$, while column eight contains
the values of $\eta=2-\gamma/\nu$. The three sets of parameters corresponding
to the same $L_{\rm min}$ refer to the cases of subleading
corrections (term with the exponent~$\delta$ in Eqs.~(\ref{fit_def}))
included (i) in neither fit, (ii) only in the fit for $\gamma/\nu$,
(iii) in both fits.}
\begin{center}
\setlength{\tabcolsep}{4.5pt}
\begin{tabular}{|c|c|c|c|c|c|c|c|}
\hline
$N$ & $L_{\min}$ & $\beta/\nu$ & $\chi^2_{\beta/\nu}$ & $\gamma/\nu$ 
& $\chi^2_{\gamma/\nu}$ & $d$ & $\eta$ \\
\hline
 2 & 8  & 0.5065(3) & 1.32 & 2.002(2)   & 26.1 & 3.015(3)  & -0.002(2)  \\
   &    & 0.5065(3) & 1.32 & 1.9608(17) & 0.73 & 2.974(2)  & 0.0392(17) \\
   &    & 0.5074(16)& 1.38 & 1.9608(17) & 0.73 & 2.976(5)  & 0.0392(17) \\
\cline{2-8}
   & 16 & 0.5066(6) & 1.50 & 1.9908(14) & 2.91 & 3.004(2)  & 0.0092(14) \\
   &    & 0.5066(6) & 1.50 & 1.962(4)   & 0.58 & 2.976(5)  & 0.038(4)   \\
   &    & 0.515(3)  & 1.08 & 1.962(4)   & 0.58 & 2.993(11) & 0.038(4)   \\
\cline{2-8}
   & 24 & 0.5058(9) & 0.87 & 1.9851(14) & 0.90 & 2.997(3)  & 0.0149(14) \\
   &    & 0.5058(9) & 0.87 & 1.971(11)  & 0.82 & 2.982(12) & 0.029(11)  \\
   &    & 0.516(7)  & 0.75 & 1.971(11)  & 0.82 & 3.00(2)   & 0.029(11)  \\
\hline
 4 & 8  & 0.5022(6) & 7.05 & 2.007(3)   & 23.5 & 3.0012(4) & -0.007(3)  \\
   &    & 0.5022(6) & 7.05 & 1.954(2)   & 0.69 & 2.958(3)  & 0.046(2)   \\
   &    & 0.5100(19)& 3.36 & 1.954(2)   & 0.69 & 2.974(6)  & 0.046(2)   \\
\cline{2-8}
   & 16 & 0.5020(7) & 7.46 & 2.002(2)   & 12.2 & 3.006(4)  & -0.002(2)  \\
   &    & 0.5020(7) & 7.46 & 1.955(2)   & 0.70 & 2.959(4)  & 0.045(2)   \\
   &    & 0.514(2)  & 2.37 & 1.955(2)   & 0.70 & 2.982(7)  & 0.045(2)   \\
\cline{2-8}
   & 24 & 0.5018(9) & 7.47 & 1.998(2)   & 6.95 & 3.001(4)  & 0.002(2)   \\
   &    & 0.5018(9) & 7.47 & 1.955(3)   & 0.76 & 2.959(5)  & 0.045(3)   \\
   &    & 0.517(2)  & 1.69 & 1.955(3)   & 0.76 & 2.989(8)  & 0.045(3)   \\
\hline
 5 & 8  & 0.5106(2) & 3.34 & 1.998(3)   & 29.8 & 3.019(3)  & 0.002(3)   \\
   &    & 0.5106(2) & 3.34 & 1.954(2)   & 0.90 & 2.975(2)  & 0.046(2)   \\
   &    & 0.5091(12)& 3.16 & 1.954(2)   & 0.90 & 2.972(4)  & 0.046(2)   \\
\cline{2-8}
   & 16 & 0.5113(3) & 1.85 & 1.9857(14) & 2.27 & 3.008(2)  & 0.0143(14) \\
   &    & 0.5113(3) & 1.85 & 1.963(6)   & 0.94 & 2.986(7)  & 0.037(6)   \\
   &    & 0.5157(17)& 1.11 & 1.963(6)   & 0.94 & 2.995(9)  & 0.037(6)   \\
\cline{2-8}
   & 24 & 0.5101(3) & 0.62 & 1.982(2)   & 1.81 & 3.002(2)  & 0.018(2)   \\
   &    & 0.5101(3) & 0.62 & 1.959(16)  & 1.52 & 2.980(17) & 0.041(16)  \\
   &    & 0.512(3)  & 0.74 & 1.959(16)  & 1.52 & 2.98(2)   & 0.041(16)  \\
\hline
 6 & 8  & 0.5077(2) & 3.36 & 2.007(4)   & 47.5 & 3.022(4)  & -0.007(4)  \\
   &    & 0.5077(2) & 3.36 & 1.949(2)   & 0.98 & 2.964(2)  & 0.051(2)   \\
   &    & 0.5071(12)& 3.59 & 1.949(2)   & 0.98 & 2.963(4)  & 0.051(2)   \\
\cline{2-8}
   & 16 & 0.5078(2) & 0.88 & 1.990(2)   & 4.63 & 3.006(2)  & 0.010(2)   \\
   &    & 0.5078(2) & 0.88 & 1.947(7)   & 0.84 & 2.963(7)  & 0.053(7)   \\
   &    & 0.5121(12)& 0.34 & 1.947(7)   & 0.84 & 2.971(9)  & 0.053(7)   \\
\cline{2-8}
   & 24 & 0.5070(3) & 0.45 & 1.983(2)   & 1.76 & 2.997(3)  & 0.017(2)   \\
   &    & 0.5070(3) & 0.45 & 1.95(3)    & 1.85 & 2.97(3)   & 0.05(3)    \\
   &    & 0.510(4)  & 0.58 & 1.95(3)    & 1.85 & 2.97(4)   & 0.05(3)    \\
\hline
\end{tabular}
\end{center}
\label{indices_part1}
\end{table}

\begin{table}
\caption{Same as Table~\ref{indices_part1} for $3D$ $Z(N)$ vector models 
with $N=8,13,20.$}
\begin{center}
\setlength{\tabcolsep}{4.5pt}
\begin{tabular}{|c|c|c|c|c|c|c|c|}
\hline
$N$ & $L_{\min}$ & $\beta/\nu$ & $\chi^2_{\beta/\nu}$ & $\gamma/\nu$ 
& $\chi^2_{\gamma/\nu}$ & $d$ & $\eta$ \\
\hline
 8 & 8  & 0.5083(2) & 4.00 & 2.006(3)   & 66.0 & 3.023(3)  & -0.006(4)  \\
   &    & 0.5083(2) & 4.00 & 1.952(2)   & 1.57 & 2.968(2)  & 0.048(2)   \\
   &    & 0.5082(9) & 4.28 & 1.952(2)   & 1.57 & 2.968(4)  & 0.048(2)   \\
\cline{2-8}
   & 16 & 0.5085(3) & 2.87 & 1.992(2)   & 10.5 & 3.009(2)  & 0.008(2)   \\
   &    & 0.5085(3) & 2.87 & 1.947(6)   & 2.04 & 2.964(6)  & 0.053(6)   \\
   &    & 0.5136(13)& 1.33 & 1.947(6)   & 2.04 & 2.974(9)  & 0.053(6)   \\
\cline{2-8}
   & 24 & 0.5079(5) & 3.52 & 1.983(2)   & 4.57 & 2.999(3)  & 0.017(2)   \\
   &    & 0.5079(5) & 3.52 & 1.944(16)  & 2.86 & 2.959(17) & 0.056(16)  \\
   &    & 0.519(3)  & 1.43 & 1.944(16)  & 2.86 & 2.98(2)   & 0.056(16)  \\
\hline
 13& 8  & 0.5087(3) & 8.22 & 2.011(3)   & 52.5 & 3.028(4)   & -0.011(3) \\
   &    & 0.5087(3) & 8.22 & 1.956(2)   & 1.45 & 2.973(3)   & 0.044(2)  \\  
   &    & 0.5055(13)& 6.28 & 1.956(2)   & 1.45 & 2.967(5)   & 0.044(2)  \\
\cline{2-8}
   & 16 & 0.5097(3) & 2.05 & 1.994(2)   & 6.87 & 3.014(2)   & 0.006(2)  \\  
   &    & 0.5097(3) & 2.05 & 1.949(6)   & 1.48 & 2.969(7)   & 0.051(6)  \\ 
   &    & 0.5137(17)& 1.48 & 1.949(6)   & 1.48 & 2.977(10)  & 0.051(6)  \\  
\cline{2-8}
   & 24 & 0.5088(5) & 2.12 & 1.985(2)   & 2.06 & 3.003(3)   & 0.015(2)  \\  
   &    & 0.5088(5) & 2.12 & 1.958(15)  & 1.61 & 2.976(16)  & 0.042(15) \\  
   &    & 0.514(4)  & 2.10 & 1.958(15)  & 1.61 & 2.99(2)    & 0.042(15) \\  
\hline
 20& 8  & 0.5076(3) & 9.69 & 2.006(3)   & 65.7 & 3.022(4)   & -0.006(3) \\  
   &    & 0.5076(3) & 9.69 & 1.952(2)   & 1.94 & 2.967(2)   & 0.048(2)  \\  
   &    & 0.5067(14)& 10.0 & 1.952(2)   & 1.94 & 2.965(5)   & 0.048(2)  \\  
\cline{2-8}
   & 16 & 0.5080(3) & 4.48 & 1.9905(18) & 7.25 & 3.007(2)   & 0.0095(18)\\  
   &    & 0.5080(3) & 4.48 & 1.953(5)   & 1.82 & 2.969(6)   & 0.047(5)  \\  
   &    & 0.5148(16)& 1.92 & 1.953(5)   & 1.82 & 2.983(9)   & 0.047(5)  \\  
\cline{2-8}
   & 24 & 0.5074(7) & 5.80 & 1.985(2)   & 4.68 & 2.999(4)   & 0.015(2)  \\  
   &    & 0.5074(7) & 5.80 & 1.941(13)  & 2.33 & 2.956(15)  & 0.059(13) \\  
   &    & 0.523(2)  & 0.85 & 1.941(13)  & 2.33 & 2.987(18)  & 0.059(13) \\  
\hline
\end{tabular}
\end{center}
\label{indices_part2}
\end{table}

\begin{table}
\caption{Critical indices of $3D$ $Z(N)$ vector models obtained from 
two alternative methods (see the text); the results of the first alternative
are given in column two, those of the second alternative (three variants)
in columns from three to five.
The {\it legenda} for the entries in each large cell is the following: 
$\eta=2-\gamma/\nu$ (first line), $\beta/\nu$ with its $\chi^2$ (second line), 
$\gamma/\nu$ with its $\chi^2$ (third line) and the dimension 
$d=2\gamma/\nu+\beta/\nu$ (fourth line). Subleading corrections (term with 
the exponent~$\delta$) were not included in either fits.}
\begin{center}
\setlength{\tabcolsep}{4.5pt}
\begin{tabular}{|c|c|c|c|c|}
\hline
$N$ & $\beta_{\rm pc}$, $\chi^{(M)}_L$ max & $\beta_{\rm f}, {U_L^{(M)}}_{\rm f} = {U_L^{(M)}}_{\rm crit}$ 
& $\beta_{\rm f}, {U_L^{(M)}}_{\rm f} > {U_L^{(M)}}_{\rm f}$ & $\beta_{\rm f}, {U_L^{(M)}}_{\rm f} < {U_L^{(M)}}_{\rm f}$ \\ 
\hline
 2 & $\eta$=0.017(2) & 0.0172(13) & 0.0162(19) & 0.0141(15)  \\
   & $\beta/\nu$=0.512(3), $\chi^2$=6.18 & 0.504(2) 1.93 & 0.504(2) 1.93 & 0.5049(12) 1.80 \\
   & $\gamma/\nu$=1.983(2), $\chi^2$=1.53 & 1.9828(13) 0.55 & 1.9838(19) 1.48 & 1.9859(15) 2.02 \\
   & $d$=3.006(9)    & 2.990(5)   & 2.991(5)   & 2.996(4)   \\
\hline
 4 & 0.0106(18)      & 0.0182(18)      & 0.018(3)      & 0.016(2)      \\
   & 0.4983(17) 43.3 & 0.493(2)   36.6 & 0.493(2) 36.6 & 0.493(2) 36.6 \\
   & 1.9894(18) 4.28 & 1.9818(18) 1.09 & 1.982(2) 2.54 & 1.984(2) 2.49 \\
   & 2.986(5)        & 2.968(7)        & 2.967(8)      & 2.970(7)      \\
\hline
5  & 0.0218(13)      & 0.0210(10)      & 0.0219(14)      & 0.0218(13)      \\
   & 0.5106(9)  3.56 & 0.5103(4)  1.66 & 0.5088(7)  3.61 & 0.5106(9)  3.56 \\
   & 1.9762(13) 0.76 & 1.9790(10) 0.76 & 1.9781(14) 0.75 & 1.9782(13) 0.76 \\
   & 2.999(3)        & 2.9996(18)      & 2.996(2)        & 2.999(3)        \\
\hline
6  & 0.0227(16)      & 0.0179(14)      & 0.0182(12)      & 0.021(2)        \\
   & 0.5052(9)  6.04 & 0.5052(9)  6.04 & 0.5052(9)  6.04 & 0.5101(13) 3.49 \\
   & 1.9773(16) 1.92 & 1.9821(14) 1.06 & 1.9818(12) 1.37 & 1.979(2)   0.85 \\
   & 2.988(3)        & 2.992(3)        & 2.992(3)        & 2.999(4)        \\
\hline
8  & 0.0213(15)      & 0.0176(16)      & 0.0172(13)      & 0.0222(19)      \\
   & 0.511(4)   180. & 0.5075(6)  4.12 & 0.5083(8)  14.1 & 0.5098(4)  0.73 \\
   & 1.9787(15) 1.06 & 1.9824(16) 1.91 & 1.9828(13) 1.74 & 1.9778(19) 1.12 \\
   & 3.000(11)       & 2.997(2)        & 2.999(2)        & 2.998(2)        \\
\hline
13 & 0.0199(12)      & 0.0168(16)      & 0.0146(14)      & 0.0149(12)      \\
   & 0.5078(9)  3.72 & 0.5075(8)  4.16 & 0.5092(11) 20.0 & 0.5063(8)  10.1 \\
   & 1.9801(12) 0.90 & 1.9832(16) 1.24 & 1.9854(14) 1.73 & 1.9851(12) 2.07 \\
   & 2.996(3)        & 2.998(3)        & 3.004(3)        & 2.998(3)        \\
\hline
20 & 0.0221(18)      & 0.0166(13)      & 0.005(9)       & 0.0196(12)      \\
   & 0.5062(15) 11.8 & 0.5071(5)  3.43 & 0.5034(7) 16.1 & 0.5094(9)  10.7 \\
   & 1.9779(18) 2.35 & 1.9834(13) 1.26 & 1.995(9)  123. & 1.9804(12) 2.00 \\
   & 2.990(4)        & 2.998(3)        & 3.002(10)      & 2.999(3)        \\
\hline
\end{tabular}
\end{center}
\label{indices_0_method}
\end{table}

\begin{table}
\caption{Same as Table~\ref{indices_0_method}, with subleading corrections 
(term with the exponent~$\delta$) included only in the fits for $\gamma/\nu$.}
\begin{center}
\setlength{\tabcolsep}{4.5pt}
\begin{tabular}{|c|c|c|c|c|}
\hline
$N$ & $\beta_{\rm pc}$, $\chi_{M_L}$ max & $\beta_{\rm f}, {U_L^{(M)}}_{\rm f} = {U_L^{(M)}}_{\rm crit}$ 
& $\beta_{\rm f}, {U_L^{(M)}}_{\rm f} > {U_L^{(M)}}_{\rm f}$ & $\beta_{\rm f}, {U_L^{(M)}}_{\rm f} < {U_L^{(M)}}_{\rm f}$ \\ 
\hline
 2 & 0.03(2)       & 0.032(11)      & 0.025(18)      & 0.033(10)       \\
   & 0.512(3) 6.18 & 0.504(2)  1.93 & 0.504(2)  1.93 & 0.5049(12) 1.80 \\
   & 1.97(2)  1.74 & 1.968(11) 0.49 & 1.975(8)  1.67 & 1.967(10)  1.68 \\
   & 3.00(3)       & 2.975(15)      & 2.98(2)        & 2.977(13)       \\
\hline
 4 & 0.041(5)        & 0.039(12)      & 0.049(14)      & 0.044(6)      \\
   & 0.4983(17) 43.3 & 0.493(2)  36.6 & 0.493(2)  36.6 & 0.493(2) 36.6 \\
   & 1.959(5)   1.15 & 1.961(12) 0.89 & 1.951(14) 1.74 & 1.956(6) 0.80 \\
   & 2.956(8)        & 2.947(17)      & 2.936(19)      & 2.942(11)     \\
\hline
5  & 0.036(8)        & 0.037(5)       & 0.029(12)      & 0.036(8)       \\
   & 0.5106(9)  3.56 & 0.5103(4) 1.66 & 0.5088(7) 3.61 & 0.5106(9) 3.56 \\
   & 1.964(8)   0.67 & 1.963(5)  0.44 & 1.971(12) 0.80 & 1.964(8)  0.67 \\
   & 2.985(10)       & 2.984(5)       & 2.989(13)      & 2.985(10)      \\
\hline
6  & 0.040(11)      & 0.038(10)      & 0.030(5)       & 0.026(16)       \\
   & 0.5052(9) 6.04 & 0.5052(9) 6.04 & 0.5052(9) 6.04 & 0.5101(13) 3.49 \\
   & 1.960(11) 1.68 & 1.962(10) 0.77 & 1.970(5)  0.95 & 1.974(16)  0.97 \\
   & 2.970(13)      & 2.972(11)      & 2.980(7)       & 2.994(19)       \\
\hline
8  & 0.038(13)      & 0.042(10)      & 0.034(4)       & 0.048(12)      \\
   & 0.511(4)  180. & 0.5075(6) 4.12 & 0.5083(8) 14.1 & 0.5098(4) 0.73 \\
   & 1.962(13) 1.00 & 1.958(10) 1.27 & 1.966(4)  0.65 & 1.952(12) 0.82 \\
   & 2.98(2)        & 2.973(11)      & 2.983(6)       & 2.972(13)      \\
\hline
13  & 0.031(10)      & 0.038(9)       & 0.029(7)        & 0.038(6)       \\
    & 0.5078(9) 3.72 & 0.5075(8) 4.16 & 0.5092(11) 20.0 & 0.5063(8) 10.1 \\
    & 1.969(10) 0.89 & 1.962(9)  0.87 & 1.971(7)   1.37 & 1.962(6)  1.12 \\
    & 2.984(12)      & 2.977(11)      & 2.990(10)       & 2.975(8)       \\
\hline
20  & 0.036(10)       & 0.036(8)       & 0.01(6)        & 0.033(4)       \\
    & 0.5062(15) 11.8 & 0.5071(5) 3.43 & 0.5034(7) 16.1 & 0.5094(9) 10.7 \\
    & 1.964(10)  2.10 & 1.964(8)  0.84 & 1.99(6)   134. & 1.967(4)  1.00 \\
    & 2.976(13)       & 2.978(9)       & 3.00(6)        & 2.985(6)       \\
\hline
\end{tabular}
\end{center}
\label{indices_1_method}
\end{table}

\begin{table}
\caption{Same as Table~\ref{indices_0_method}, with subleading corrections 
(term with the exponent~$\delta$) included in both the fits for $\gamma/\nu$ 
and for $\beta/\nu$.}
\begin{center}
\setlength{\tabcolsep}{4.5pt}
\begin{tabular}{|c|c|c|c|c|}
\hline
$N$ & $\beta_{\rm pc}$, $\chi_{M_L}$ max &$\beta_{\rm f}, {U_L^{(M)}}_{\rm f} = {U_L^{(M)}}_{\rm crit}$ 
& $\beta_{\rm f}, {U_L^{(M)}}_{\rm f} > {U_L^{(M)}}_{\rm f}$ & $\beta_{\rm f}, {U_L^{(M)}}_{\rm f} < {U_L^{(M)}}_{\rm f}$ \\ 
\hline
 2 & 0.03(2)        & 0.032(11)      & 0.025(18)     & 0.033(10)      \\
   & 0.534(11) 1.37 & 0.536(8)  0.63 & 0.536(8) 0.63 & 0.524(5)  0.78 \\
   & 1.97(2)   1.74 & 1.968(11) 0.49 & 1.975(8) 1.67 & 1.967(10) 1.68 \\
   & 3.04(4)        & 3.04(2)        & 3.05(3)       & 3.01(2)        \\
\hline
 4 & 0.041(5)      & 0.039(12)      & 0.049(14)      & 0.044(6)      \\
   & 0.527(2) 3.51 & 0.535(3)  1.71 & 0.535(3)  1.71 & 0.535(3) 1.71 \\
   & 1.959(5) 1.15 & 1.961(12) 0.89 & 1.951(14) 1.74 & 1.956(6) 0.80 \\
   & 3.012(10)     & 3.032(19)      & 3.02(2)        & 3.027(13)     \\
\hline
5  & 0.036(8)      & 0.037(5)        & 0.029(12)      & 0.036(8)      \\
   & 0.504(6) 3.48 & 0.5035(13) 0.52 & 0.500(4)  3.04 & 0.504(6) 3.48 \\
   & 1.964(8) 0.67 & 1.963(5)   0.44 & 1.971(12) 0.80 & 1.964(8) 0.67 \\
   & 2.97(2)       & 2.970(7)        & 2.97(2)        & 2.97(2)       \\
\hline
6  & 0.040(11)      & 0.038(10)      & 0.030(5)      & 0.026(16)       \\
   & 0.523(2)  0.97 & 0.523(2)  0.97 & 0.523(2) 0.97 & 0.488(4)   0.98 \\
   & 1.960(11) 1.68 & 1.962(10) 0.77 & 1.970(5) 0.95 & 1.974(16)  0.97 \\
   & 3.005(17)      & 3.007(15)      & 3.015(11)     & 2.95(2)         \\
\hline
8  & 0.038(13)      & 0.042(10)      & 0.034(4)      & 0.048(12)      \\
   & 0.58(4)   149. & 0.497(3)  1.72 & 0.497(2) 3.43 & 0.509(3)  0.82 \\
   & 1.962(13) 1.00 & 1.958(10) 1.27 & 1.966(4) 0.65 & 1.952(12) 0.82 \\
   & 3.11(10)       & 2.952(16)      & 2.960(8)      & 2.97(2)        \\
\hline
13  & 0.031(10)      & 0.038(9)      & 0.029(7)        & 0.038(6)      \\
    & 0.508(6)  4.25 & 0.494(2) 0.95 & 0.4924(17) 1.56 & 0.493(4) 5.99 \\
    & 1.969(10) 0.89 & 1.962(9) 0.87 & 1.971(7)   1.37 & 1.962(6) 1.12 \\
    & 2.98(2)        & 2.951(14)     & 2.956(11)       & 2.948(15)     \\
\hline
20  & 0.036(10)      & 0.036(8)      & 0.01(6)       & 0.033(4)        \\
    & 0.503(9)  13.6 & 0.497(2) 1.15 & 0.491(2) 4.35 & 0.4976(19) 1.99 \\
    & 1.964(10) 2.10 & 1.964(8) 0.84 & 1.99(6)  134. & 1.967(4)   1.00 \\
    & 2.97(2)        & 2.957(13)     & 2.98(7)       & 2.962(8)        \\
\hline
\end{tabular}
\end{center}
\label{indices_2_method}
\end{table}

\begin{figure}
\centering
\includegraphics[width=0.49\textwidth]{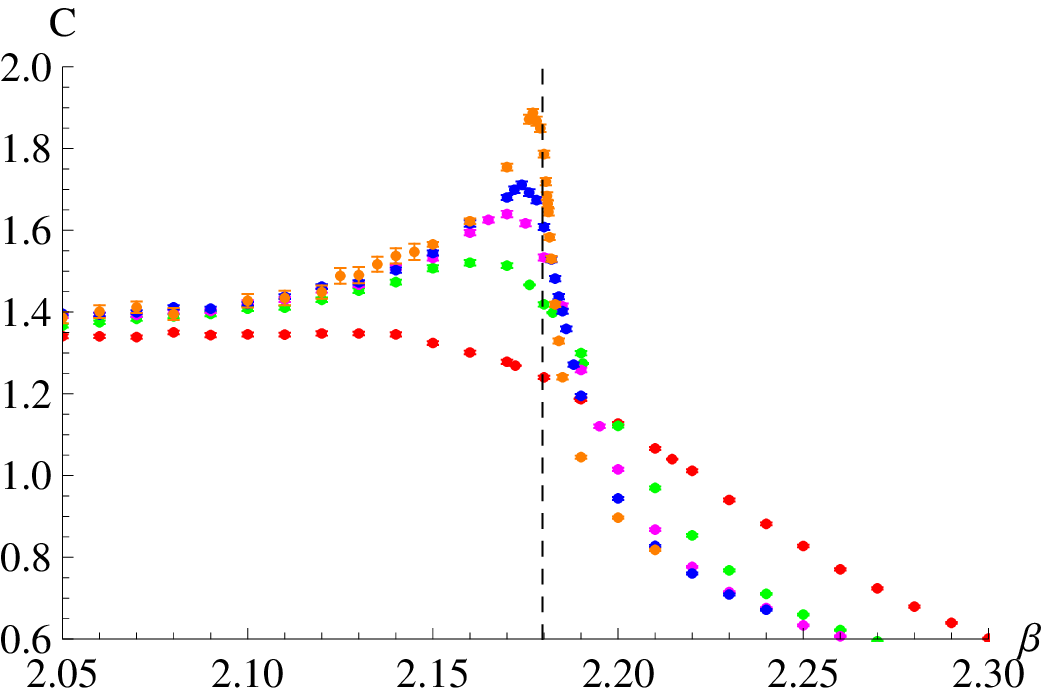}
\includegraphics[width=0.49\textwidth]{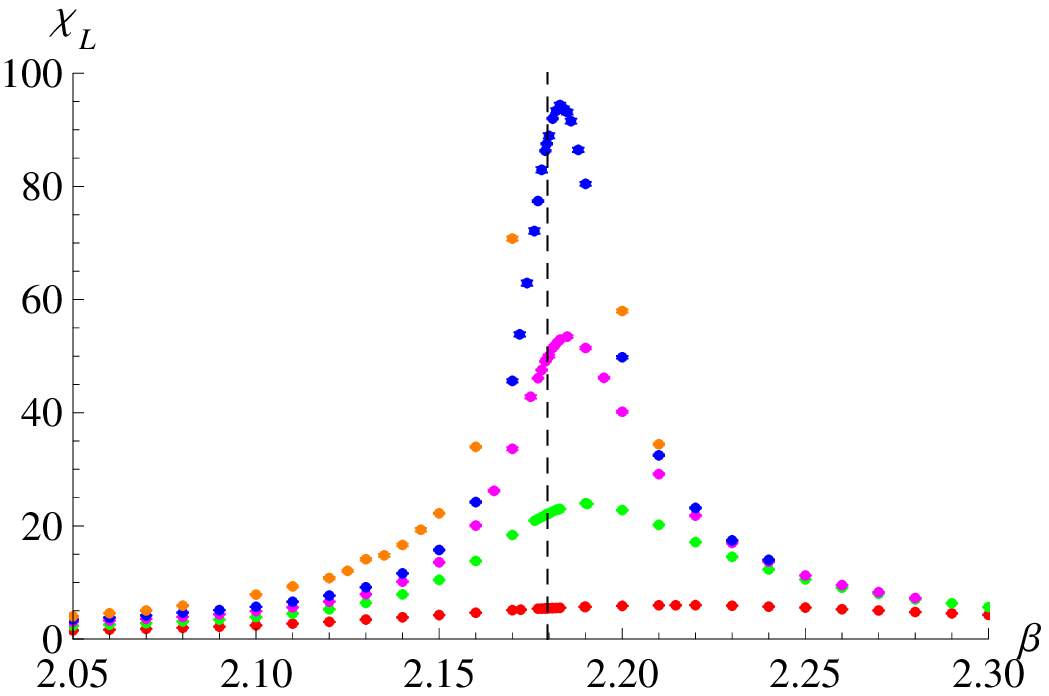}
\caption{Heat capacity (left) and susceptibility $\chi_L^{(M)}$ (right) 
{\it versus} $\beta$ for the $3D$ $Z(5)$ vector model on lattices with 
$L$ = 8 (red), 16 (green), 24 (violet), 32 (blue) and 64 (orange). The 
vertical dashed line shows the location of the critical point. The error bars
are smaller than symbols.}
\label{spchiplots}
\end{figure}

\subsection{Heat capacity and the index $\alpha$} 

The critical index $\alpha$, determined from the $\nu$ values obtained
in the previous subsection by means of the relation $\alpha=2-d\nu$, gets
negative values for all $N\geq 5$, meaning that the transition is of order higher than two. 
In fact, these negative values are very close to that of the $3D$ $XY$ model \cite{3dxy_univ}.  
However, the plots of the heat capacity (see Fig.~\ref{spchiplots}(left)) 
clearly show that it diverges in the vicinity of the critical point. 
Moreover, the maxima of the heat capacity and of the susceptibility $\chi_L^{(M)}$ 
approach the critical point from different sides (see Figs.~\ref{spchiplots}).

This suggests that a different value for the index $\nu$ can be found
if a FSS analysis is done on the peak values of the heat capacity,
using as fitting function
\begin{equation}
C(L) = A L^{\alpha/\nu} (1 + B L^{-\delta} )\ ,
\label{hc_fit}
\end{equation}
where the possible inclusion of subleading corrections has been taken into
account. After $\alpha/\nu$ is extracted, from the relation $\alpha=2-d \nu$
the value of $\nu$ can be obtained.

In Table~\ref{indices_nu_zero_hc} we summarize the results for $\nu$
determined in the described way, with and without the inclusion of 
the subleading correction term. When considered, the exponent $\delta$ 
has been fixed to the value $0.5/\nu$. 
The error estimates given in the table do not include systematic uncertainties
brought by the localization of the maximum of the heat capacity by
using analytic continuation, which can be unreliable in regions where 
the heat capacity changes fast. The constant $B$ in front of the subleading 
correction appears to be of order unity, $B\sim 0.4-0.7$, and does not seem 
to depend on $N$. 

We see that while for $N=2, 4$ the resulting $\nu$ agrees with the value of 
$\nu$ in the $3D$ Ising model, and the agreement improves if we include the
subleading correction, for $N>4$ this is not the case. For $N>4$ the 
difference between the $\nu$ values obtained with and without inclusion of 
subleading corrections is much smaller than for $N=2,4$. The most important
fact is, however, that in all cases the $\nu$ indices obtained in this way
are close to $\nu \approx 0.63$ -- the critical index for the Ising model. 
The difference between $\nu$ indices obtained from the $U_L^{(M)}$ cumulants 
and from the heat capacity leads us to conclude that we have two kinds of 
singularity depending on whether one approaches the critical coupling from 
above ($3D$ $XY$ model-like singularity) or from below 
($3D$ Ising universality class), for $N > 4$. 

\begin{table}
\caption{Critical index $\nu$ of $3D$ $Z(N)$ vector models with $N=2,4,5,6,
8,13,20$ determined by the fit given in Eq.~(\ref{hc_fit}) on 
the peak of the heat capacity, for different choices of the minimum lattice 
size $L_{\rm min}$. The two sets of parameters refer to the cases of 
subleading corrections (term with the exponent~$\delta$ in Eqs.~(\ref{hc_fit}))
not included in the fit or included with $\delta$ fixed at $0.5/\nu$ 
The $\chi^2$ of the fits, given in columns four and six, are the reduced ones.}
\begin{center}
\setlength{\tabcolsep}{4.5pt}
\begin{tabular}{|c|c|c|c|c|c|}
\hline
    &            & \multicolumn{2}{|c|}{no subl. corr.} & \multicolumn{2}{|c|}{with subl. corr.} \\
\cline{3-6}
$N$ & $L_{\min}$ & $\nu$  & $\chi^2_{\nu}$ & $\nu$  & $\chi^2_{\nu}$ \\
\hline
 2 & 8  & 0.6108(4)  & 8.00 & 0.6185(4)  & 0.44 \\
   & 16 & 0.6132(4)  & 1.37 & 0.6214(16) & 0.42 \\
   & 24 & 0.6143(6)  & 1.04 & 0.623(4)   & 0.46 \\
   & 32 & 0.6154(7)  & 0.51 & 0.627(7)   & 0.42 \\
\hline
 4 & 8  & 0.6117(9)  & 3.27 & 0.6223(17) & 0.79\\
   & 16 & 0.6146(9)  & 1.16 & 0.629(4)   & 0.72\\
   & 24 & 0.6168(14) & 0.81 & 0.634(9)   & 0.57\\
\hline
 5 & 8  & 0.6047(6)  & 5.35 & 0.639(2)  & 0.89\\
   & 16 & 0.6338(12) & 1.94 & 0.6410(8) & 0.08\\
   & 24 & 0.6360(6)  & 0.22 & 0.6409(15)& 0.17\\
\hline
 6 & 8  & 0.6300(10) & 21.3 & 0.640(5) & 16.5\\
   & 16 & 0.6348(8)  & 1.78 & 0.642(3) & 0.97\\
   & 24 & 0.6360(10) & 1.57 & 0.646(3) & 0.61\\
\hline
 8 & 8  & 0.6293(6)  & 78.6 & 0.6385(18) & 19.3\\
   & 16 & 0.6320(4)  & 11.4 & 0.637(3)   & 9.71\\
   & 24 & 0.6336(3)  & 2.40 & 0.631(10)  & 11.8\\
\hline
13 & 8  & 0.6294(6)  & 68.6 & 0.6411(3)  & 2.11 \\
   & 16 & 0.6323(4)  & 9.04 & 0.6425(7)  & 1.80 \\
   & 24 & 0.6340(4)  & 2.52 & 0.6442(13) & 1.87 \\
\hline
20 & 8  & 0.6285(5)  & 53.8 & 0.6428(8)  & 6.79\\
   & 16 & 0.6304(3)  & 5.19 & 0.6472(17) & 5.19\\
   & 24 & 0.6314(3)  & 2.35 & 0.657(3)   & 2.84\\
\hline
\end{tabular}
\end{center}
\label{indices_nu_zero_hc}
\end{table}

\section{Symmetric phase}

\begin{figure}
\centering
\includegraphics[width=0.44\textwidth]{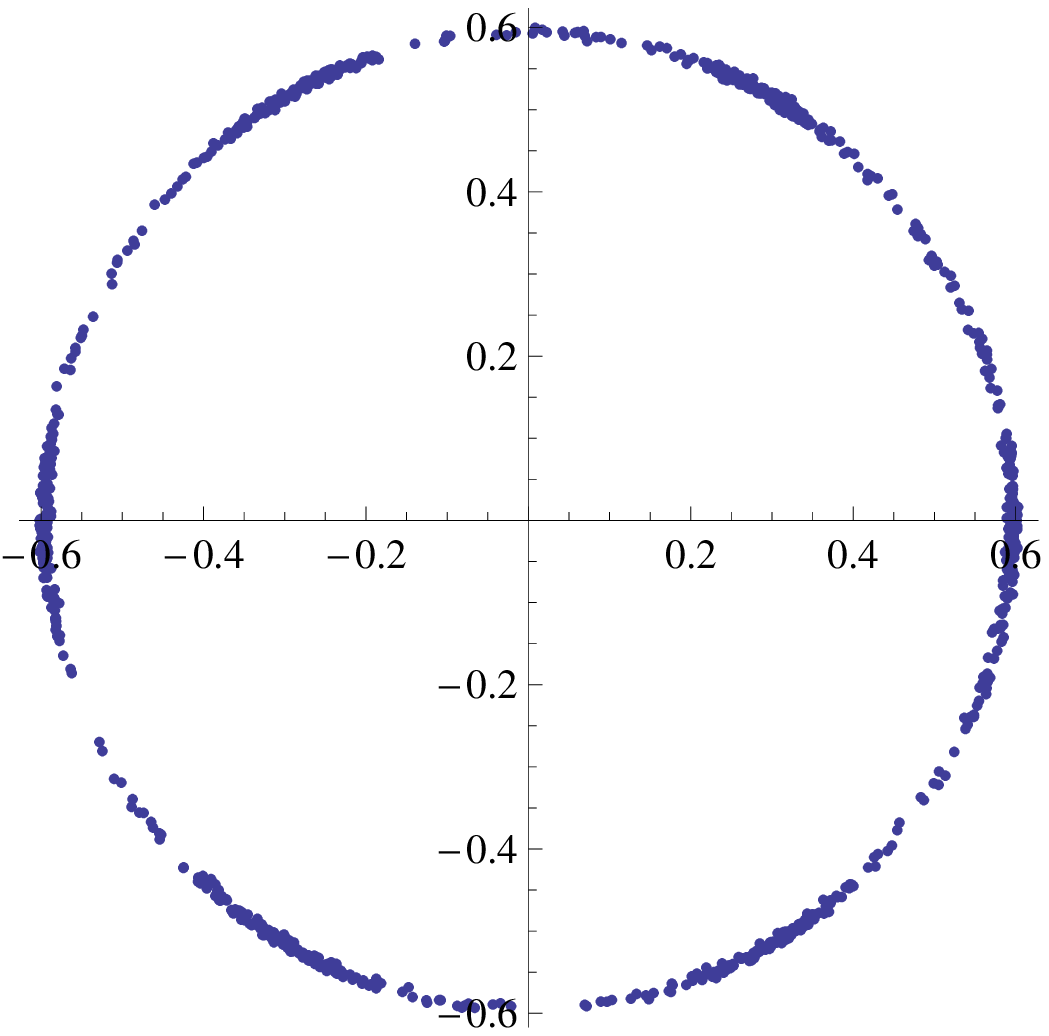}
~ 
\includegraphics[width=0.44\textwidth]{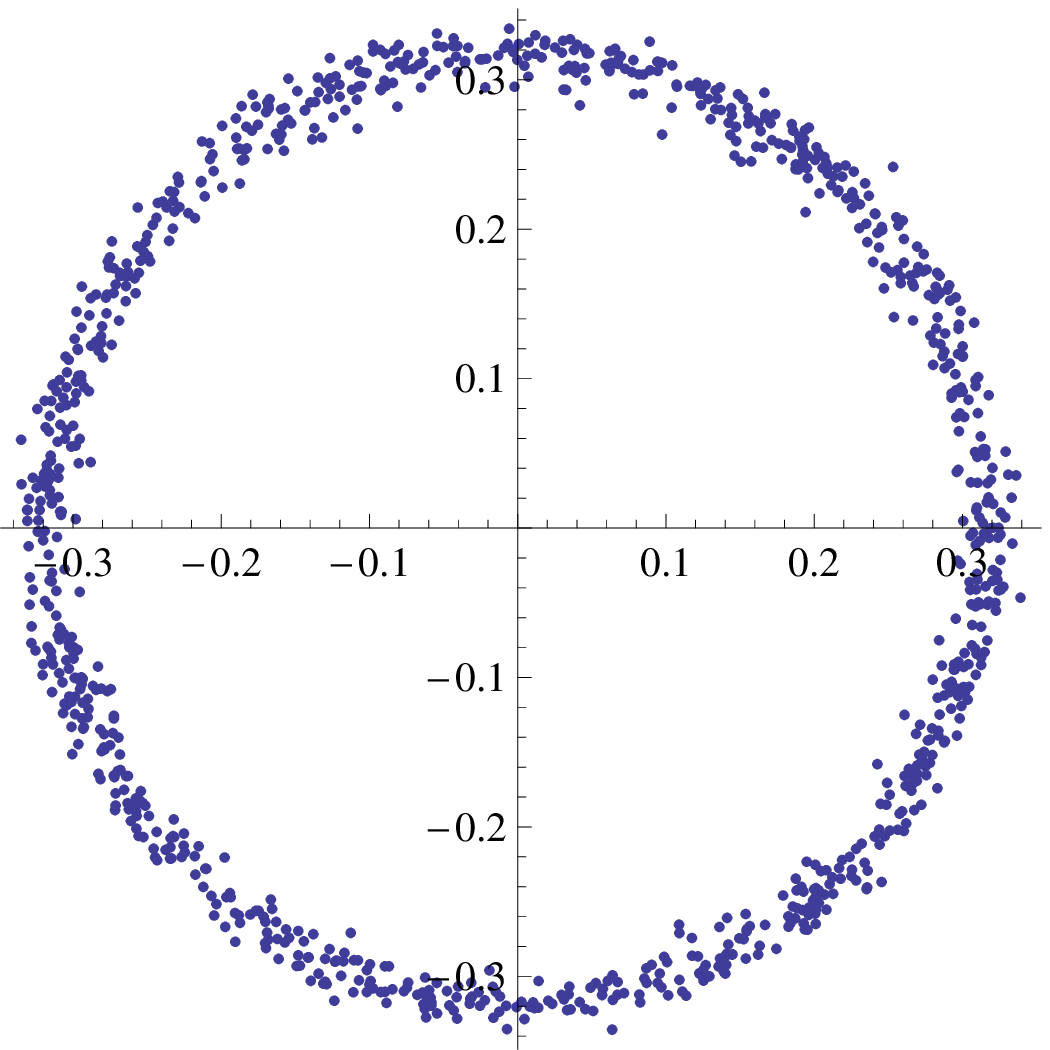}

\includegraphics[width=0.44\textwidth]{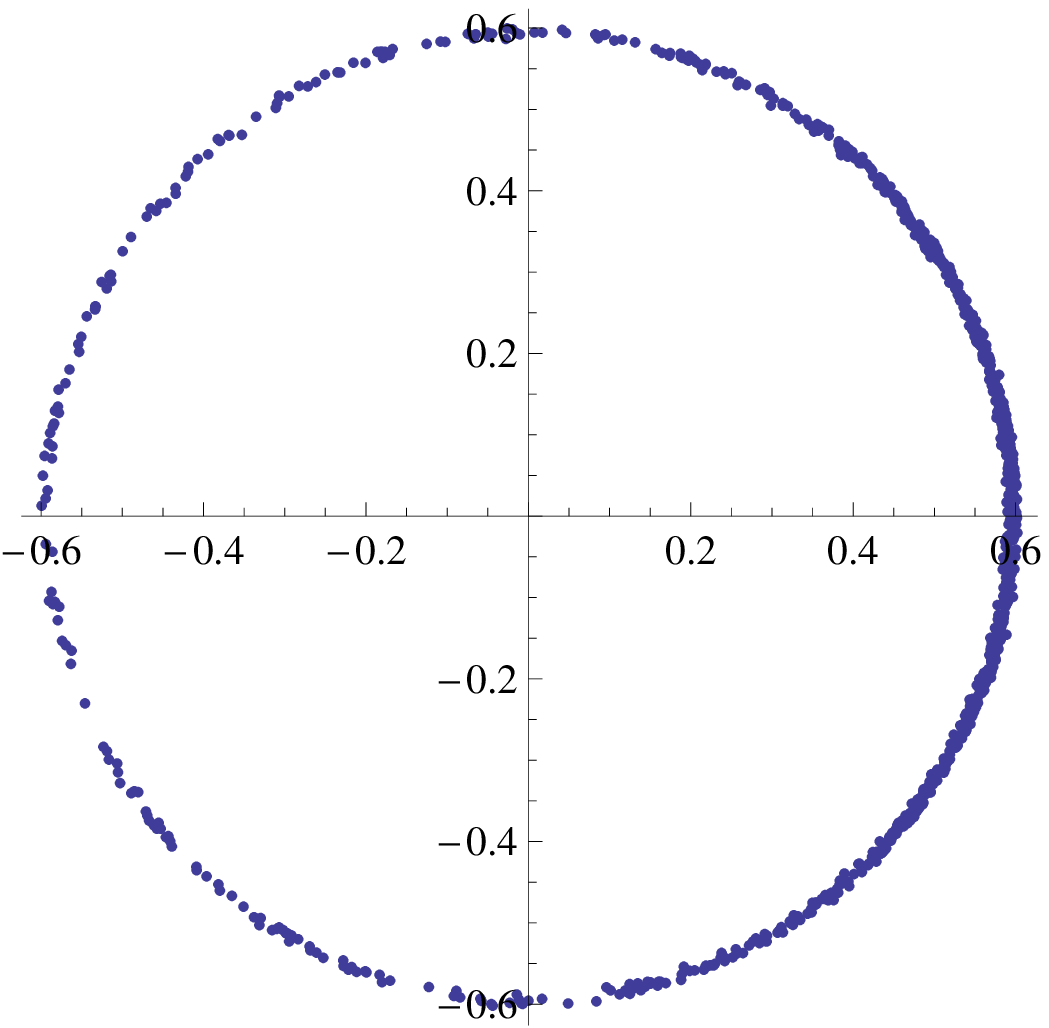}
~ 
\includegraphics[width=0.44\textwidth]{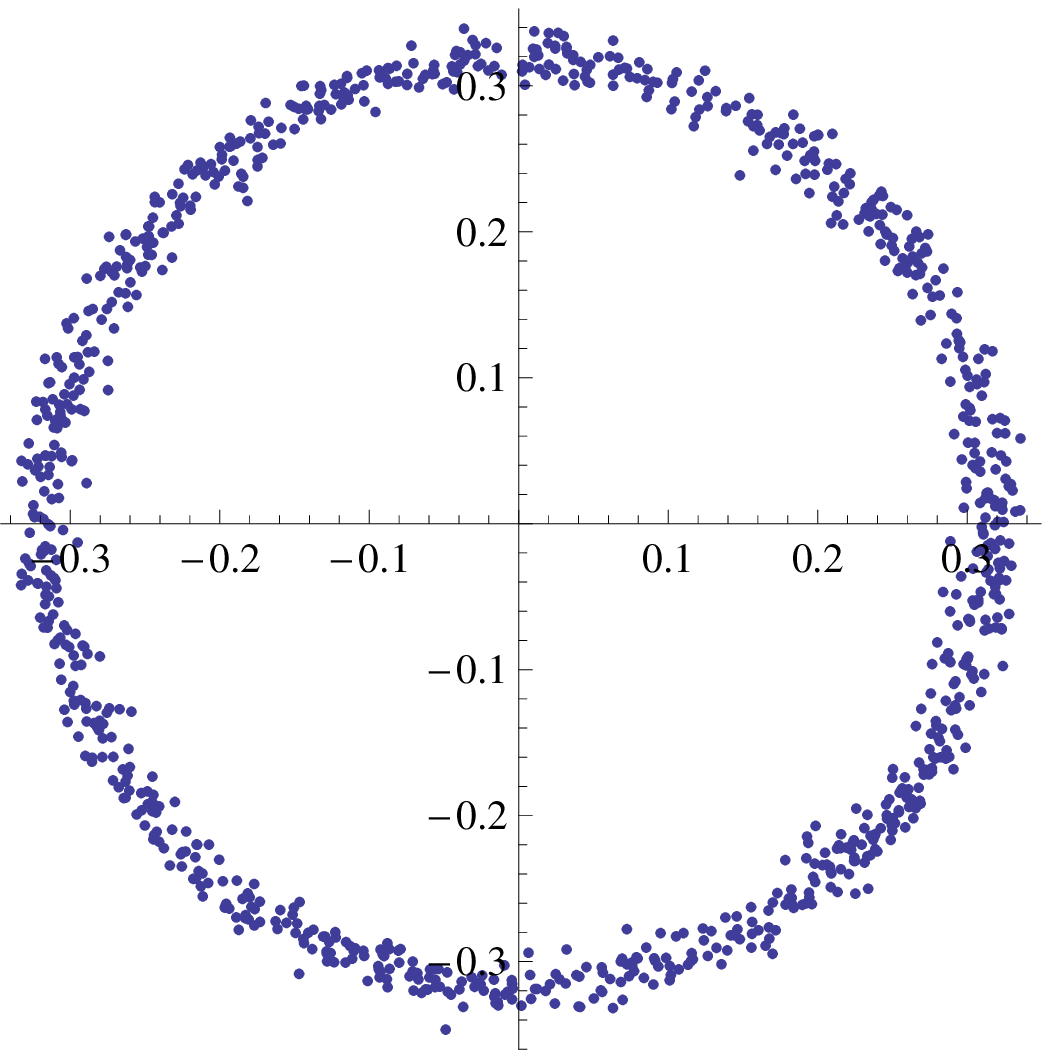}
        
\includegraphics[width=0.44\textwidth]{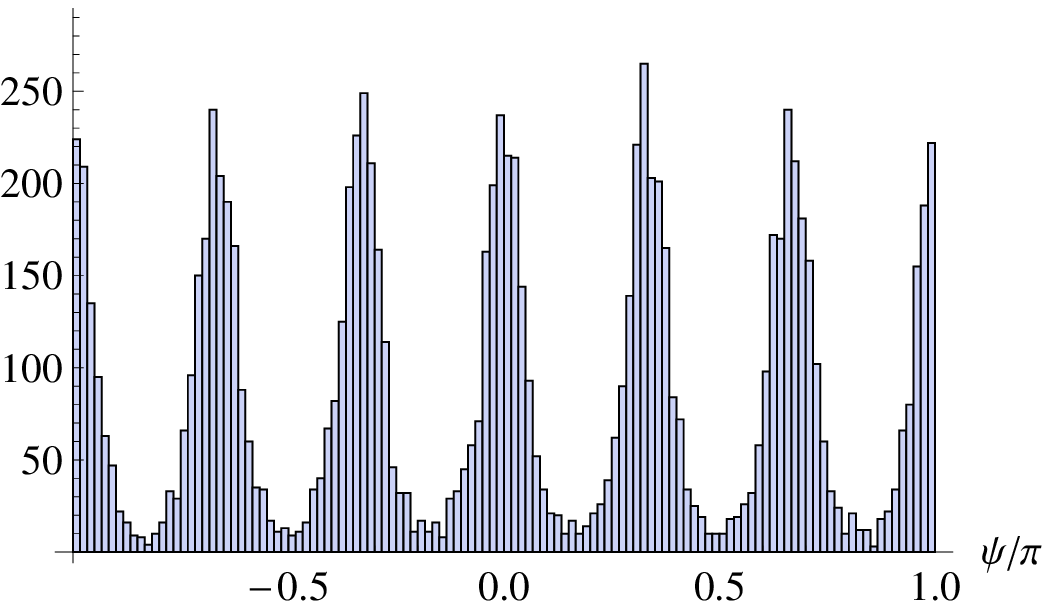}
~ 
\includegraphics[width=0.44\textwidth]{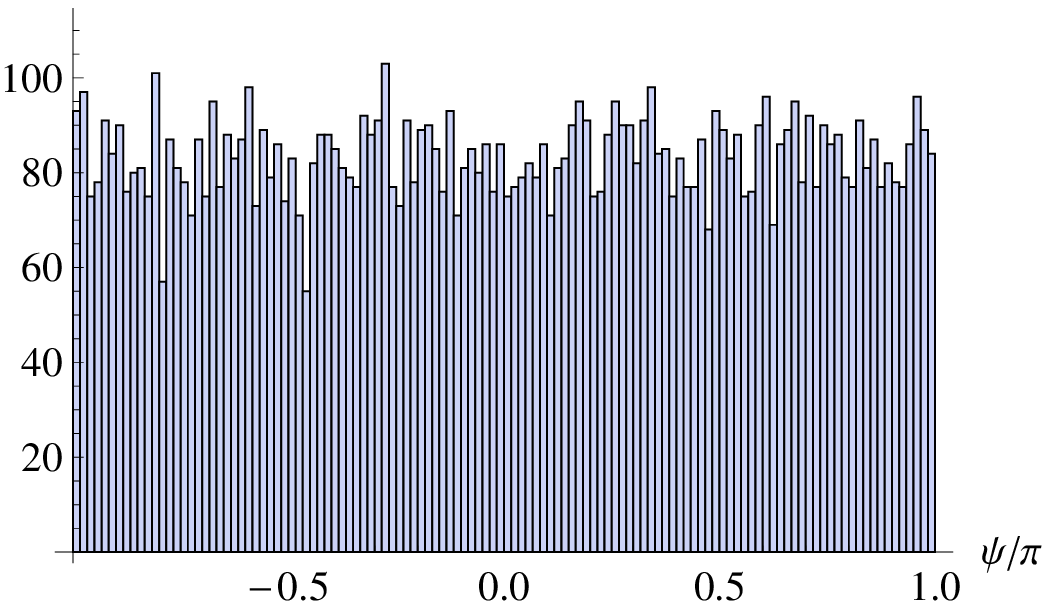}
        
\caption{Scatter plots for the magnetization (top) and the rotated 
magnetization (middle), histogram of the magnetization angle (bottom) 
for the $3D$ $Z(6)$ vector model on a $64^3$ lattice for 
$\beta = 2.7$ (left) and $\beta = 2.97$ (right).}
\label{scatter_plots}
\end{figure}

\begin{figure}
\center{\includegraphics[width=0.75\textwidth]{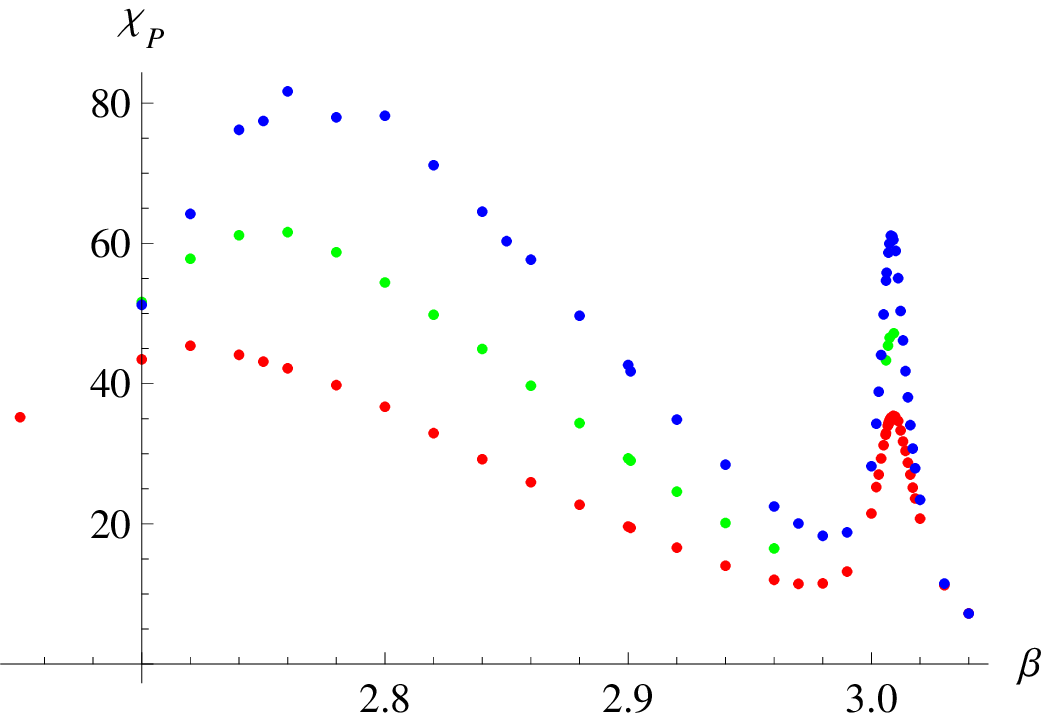}}
\caption{Population susceptibility $\chi_P$ for the $3D$ $Z(6)$ vector model 
on lattices with $L$ = 48 (red), 56 (green) and 64 (blue) {\it versus} 
$\beta$. The error bars are not shown.}
\label{chipplot}
\end{figure}

Another interesting phenomenon we have encountered during our study is the 
appearance of a symmetric phase just below $\beta_{\rm c}$ for all $N\geq 5$. 
That such phase exists in the vector $Z(6)$ spin model has been known for a 
long time~\cite{scholten,z6_vector_mc,z6_vector_rg}. 
Here we confirm its existence for all vector $Z(N\geq 5)$ LGTs. The phase 
exhibits itself, {\it e.g.}, in the behavior of magnetization. As an example, 
we give in Fig~\ref{scatter_plots} the scatter plots 
of magnetization and rotated magnetization, together with the histogram of 
the magnetization angle for $Z(6)$ on a $64^3$ lattice. 
One sees that below the critical coupling at $\beta = 2.97$ the symmetry is 
not broken on this lattice. Only starting from approximately $\beta = 2.7$ 
one can observe the appearance of a symmetry-broken phase. 
In addition, we have studied the behavior of the population susceptibility 
below $\beta_{\rm c}=3.00683$. Fig~\ref{chipplot} shows that it has a second 
broad maximum, which slowly moves to $\beta_{\rm c}=3.00683$ with increasing
lattice size. A similar picture is observed for all $N\geq 5$. 
While, however, for $N=5$ the peak of the population susceptibility moves 
rather fast and practically collapses with the peak at the critical coupling 
on the largest available lattice $L=96$, for larger $N$ the peak stays 
rather far from the corresponding critical coupling, even for $L=96$ 
(with our data we cannot even exclude a situation when the convergence of 
the second maximum is logarithmic). 
We can imagine two scenarios to explain such behavior: 
\begin{enumerate} 
\item This symmetric phase exists only in finite volume. When the lattice size 
increases, the second maximum approaches the critical coupling and, eventually,
the symmetric region shrinks and disappears. 
The explanation proposed in~\cite{z6_vector_mc,z6_vector_rg} might work in 
this case, too. 
Namely, the symmetric phase on the finite lattice is a phase with a very small 
mass gap and describes a crossover region to the symmetry-broken phase. 
\item For $N>5$ the second maximum of the population susceptibility stays 
away from the critical couplings even in the infinite volume limit. In this 
case it might correspond to some higher order phase transition 
and the symmetric phase with tiny or even vanishing mass gap exists also in 
the thermodynamic limit.
\end{enumerate}
In both cases it is tempting to speculate that this symmetric region is 
reminiscent of the massless phase which appears in these models at finite 
temperature~\cite{ZN_fin_T}. 
Whichever scenario of the above two is realized, one needs to study the 
models on much larger lattices to uncover it. 

\section{Summary} 

In this paper we have studied the $3D$ $Z(N)$ LGT at zero temperature 
aiming at shedding light on the nature of phase transitions in these models 
for $N\geq 4$. This study was based on the exact duality transformations of 
the gauge models to generalized $3D$ $Z(N)$ spin models. In Section~2 we 
presented an overview of the exact relation between couplings of these two 
models. In Section~3 we have studied the models analytically using a version 
of the phenomenological RG based on the preservation of the mass gap 
combined with a cluster decimation approximation. This study provided us with 
an approximate location of the critical couplings as well as with the value of 
the index $\nu$. 
These calculations show that the index $\nu$ is approximately the same for 
$N=2,4$, $\nu\approx 0.616$, indicating that these two models might belong to 
the same universality class. We find $\nu>2/3$ and approximately equal for 
$N>4$. Hence, $Z(N>4)$ vector LGTs belong to a different universality class. 

The numerical part of the work has been devoted to the localization of the 
critical couplings, computation of the various critical indices and check
of the hyperscaling relation. 
The main results can be shortly summarized as follows: 
\begin{itemize}
\item We have determined numerically the position of the critical couplings 
for various $Z(N)$ models. 
For $N=2,3$ we find a reasonable agreement with the values quoted in the 
literature. For larger $N$, we have significantly improved the values given 
in~\cite{bhanot}. This allowed us to improve the scaling formula for the 
critical couplings with $N$.  

\item The critical indices $\nu$ and $\eta$ derived here for $N=2,4$ suggest 
that these models are in the universality class of the $3D$ Ising model, 
while our results for all $N>4$ hint all vector $Z(N\geq 5)$ LGTs 
belong to the universality class of the $3D$ $XY$ model. This is especially 
evident from the value of the index $\nu$, which stays very close to the 
$XY$ value, $\nu\approx 0.6716$, given in~\cite{3dxy_univ}. 
The index $\alpha$ in this case takes a small negative value. It thus follows 
that a third order phase transition takes place for $N\geq 5$.    

\item A careful investigation of the specific heat and of the index $\alpha$ 
extracted from it suggests however a more complicated picture of the critical 
behavior. In this case we find a value which roughly agrees with 
the value of $3D$ Ising model for all $N$ studied. The fact that we observe 
two different values of the index $\alpha$ dependently on whether we approach 
the critical point from below or from above leads to the conclusion 
that the first derivative of the free energy could exhibit a cusp in 
the thermodynamic limit if $N>4$. 

\item Our data also revealed the existence of a symmetric phase for all $Z(N)$ 
vector LGTs if $N>4$. However, substantially larger lattices are required 
to see if this phase survives the transition to the thermodynamic limit.  

\end{itemize}

\section{Acknowledgments} 

The work of Ukrainian co-authors was supported by the Ukrainian State Fund for
Fundamental Researches under the grant F58/384-2013. Numerical simulations
have been partly carried out on Ukrainian National GRID facilities.
O.B. thanks for warm hospitality the Dipartimento di Fisica dell'Universit\`a 
della Calabria and the INFN Gruppo Collegato di Cosenza during the final 
stages of this investigation.

\end{document}